# Inorganic Crystal Structure Prototype Database based on Unsupervised Learning of Local Atomic Environments

Shulin Luo[1], Bangyu Xing[1], Muhammad Faizan[1], Jiahao Xie[1], Kun Zhou[1], Ruoting Zhao[1], Tianshu Li[1], Xinjiang Wang[2], Yuhao Fu[2,3], Xin He[1], Jian Lv[2,3], and Lijun Zhang[1,3,*]

[1]State Key Laboratory of Integrated Optoelectronics, Key Laboratory of Automobile Materials of MOE, School of Materials Science and Engineering, and Jilin Provincial International Cooperation Key Laboratory of High-Efficiency Clean Energy Materials, Jilin University, Changchun 130012, China

[2]State Key Laboratory of Superhard Materials, College of Physics, Jilin University, Changchun 130012, China

[3]International Center of Computational Method and Software, Jilin University, Changchun 130012, China

*Corresponding author: Lijun Zhang, lijun_zhang@jlu.edu.cn

**Abstract**

Recognition of structure prototypes from tremendous known inorganic crystal structures has been an important subject beneficial for material science research and new materials design. The existing databases of inorganic crystal structure prototypes were mostly constructed by classifying materials in terms of the crystallographic space group information. Herein, we employed a distinct strategy to construct the inorganic crystal structure prototype database, relying on the classification of materials in terms of local atomic environments (LAE) accompanied by unsupervised machine learning method. Specifically, we adopted a hierarchical clustering approach onto all experimentally known inorganic crystal structures data to identify structure prototypes. The criterion for hierarchical clustering is the LAE represented by the state-of-the-art structure fingerprints of the improved bond-orientational order parameters and the smooth overlap of atomic positions. This allows us to build up a LAE-based Inorganic Crystal Structure Prototype Database (LAE-ICSPD) containing 15,613 structure prototypes with defined stoichiometries. In addition, we have developed a Structure Prototype Generator Infrastructure (SPGI) package, which is a useful toolkit for structure prototype generation. Our developed SPGI toolkit and LAE-ICSPD are beneficial for investigating inorganic materials in a global way as well as accelerating materials discovery process in the data-driven mode.

# 1. Introduction

For decades, scientists have been trying to recognize structure prototypes in material databases to mark structures in a concise manner to recognize various kinds of structures[1,2]. These structure prototypes are essential for computational and experimental material research, such as high-throughput computational material design[3–8], data-driven machine learning based studies[9–11], computational crystal structure prediction[12–16], X-ray diffraction pattern analysis[17], etc. By decorating the prototype structure frames using varying combinations of atoms, scientists generate novel materials from a virtual material space[18–23]. These conjectured materials are the initial entries for computational material design frameworks[4]. Starting from the initial datasets, high-throughput computational material design frameworks first determine the thermodynamic or electronic properties using density functional theory (DFT) simulations[24–26]. By learning these structure-property data, data-driven machine learning frameworks then predict the properties of new structures with well-trained machine learning models[10]. Setting these properties as suitable screening indices, some novel and useful materials with optimal properties can be identified and analyzed. For computational crystal structure prediction, the newly generated structures are usually very similar or even identical to the original ones. Directly using these similar structures to generate the next generation will significantly slow down convergence to the global minimization solution[12,27]. Moreover, a variety of preliminary structures are essential for generating new structures with diversity[23], which enhances the global search space. Using structure prototypes to generate the preliminary structures can guarantee the uniqueness of the preliminary structures. Besides, the rules for identifying structure prototypes can be applied to identify structural similarity. Once the newly generated structures are identified (as duplicates), they will be discarded to reduce the number of calculations by orders of magnitude. On the other hand, by comparing the X-ray diffraction information of an unknown material to the corresponding information of the prototype structures, one can find the crystal structure of an unknown material. Experimentally, the structure prototypes are recorded as powder diffraction file cards. In addition, there is a positive correlation between the likelihood of success in computational discovery (e.g., advanced materials) and the data quality of the initial materials space[28]. In computational material design frameworks, using structure prototypes with high data quality for chemical substitution or structure variation is essential to design or discover advanced materials to be included in the material space.

In recent years, materials science research has taken on a new paradigm featuring data-driven science and materials informatics[11], which employs data-driven and high-throughput computational material design frameworks to discover advanced materials. In these high-throughput frameworks, by initially identifying or selecting all possible structure prototypes, materials scientists generate hypothetical compounds by placing various elements in the corresponding atomic sites of prototype structures. These virtual materials are then used as the initial inputs of the high-throughput frameworks. After DFT-based high-throughput calculations, some physical or chemical properties, such as formation energies and band gaps are estimated[4]. Based on these computations, a number of candidate materials with stable structures and optimal properties can be identified without synthesizing them first[29,30]. The high data quality and uniqueness of the structure prototypes provide greater assurance of both the high computing efficiency and success of generating target materials contained in the initial datasets of high-throughput frameworks. Therefore, structure prototypes with high data quality and uniqueness largely account for the success of high-throughput computational material design frameworks.

Structure prototypes are usually identified from experimentally known structures retrieved from existing structure databases, such as the Inorganic Crystal Structure Database (ICSD)[31,32], the Crystallography Open Database[33,34], the Cambridge Structural Database[35,36] and the Pauling File database[37]. Materials scientists mostly identify structure prototypes based on setting the space group as the first comparison. Under this rule, materials scientists have constructed several crystal structure prototype databases, such as the database embedded in the Naval Research Laboratory's *Crystal Lattice Structures* website[38], the AFLOW library of the crystallographic prototype encyclopaedia[39–41] based on the AFLOW-ICSD catalogue, and the database developed by Su et al.[42] that relies on structures retrieved from the Crystallography Open Database. All the prototype structures in these databases are identified using the space group information of crystal structures as the predominant criterion for classification. In essence, a crystal structure prototype is a typical structure that represents a set of structures. These structures share the same long-range atomic configuration type, which is an identified global atomic environment. For crystal structures, the space group is a symbolic representation of the global atomic environment, encoding only the global symmetry of a single crystal structure. Combining space group with Wyckoff positions and crystal lattice can completely describe the atomic configuration type. However, only space group is not the sole factor to describe the atomic configuration type. In fact, the global atomic environment consists of all the local atomic environments (LAEs) within one crystal structure under periodic boundary conditions. In other words, LAEs are the primary elements of global atomic environments and are identified based on the bond length, bond angle, chemical coordination information, etc. in the local region. Therefore, the atomic configuration type information is naturally encoded in the LAEs. As demonstrated in Figure S1, some crystal structures with the same space group, and even with the same Wyckoff positions, exhibit quite different LAEs, whereas some crystal structures with different space groups have identical or very similar LAEs. Moreover, the LAEs of crystal structures have a profound effect on the chemical and physical properties of materials[43–45].

With this in mind, we present a distinct approach to identify the structure prototypes by directly using LAE descriptors in conjunction with the unsupervised machine learning methods. Under our LAE-based unsupervised learning frameworks, not only are structures with very different LAEs and the same space group automatically divided into different structure prototypes, but also, structures with identical LAEs and different space groups are automatically clustered into the same structure prototype. We start from all experimentally known inorganic crystal structures retrieved from ICSD, which is the structure database with the most experimentally known inorganic crystal structures. We use state-of-the-art structure representation methods, including the bond characterization matrix (BCM)[46] and the smooth overlap of atomic positions (SOAP)[43] to fully describe the LAEs of each crystal structure. Unsupervised machine learning methods such as hierarchical clustering[47,48] are highly suitable for identifying structure prototypes because of its ability to draw boundaries between different structure prototypes and cluster the examples in the same structure prototype on large datasets with high computational efficiency[49]. We have implemented the unsupervised machine learning approach into a Structure Prototype Generator Infrastructure (SPGI) package for structure prototypes generation. These allow us to build up a LAE-based Inorganic Crystal Structure Prototype Database (LAE-ICSPD), which contains 15,613 structure porotypes with defined material stoichiometries. We finally benchmark our constructed LAE-ICSPD by calculating the cohesive energies, band gaps, and packing factors of $A_1B_2$-based structures for the hypothetical $MgCl_2$ materials.

## 2. Methods

**Structure filtering**. We retrieved all the crystallographic information file entries from the ICSD (2019.2). After removing the structures with partial occupancies of atomic sites in the unit cell, there were still some other inconsistent entries, including structures with missing atoms (e.g., H) and/or with very short interatomic separation. We first filtered out the incorrect and misfit compounds by using the *cif* module from the *ASE* package[50] to analyze and check these crystallographic information files. Then, the structures with interatomic separation below the minimum $1.2 \times r_{wigs}$ were discarded ($r_{wigs}$ stands for the Wigner Seitz radius). This rule is also adopted by the Vienna Ab initio Simulation Package (VASP)[51]. Next, the duplicate structures with the same formula symmetry were removed using a matcher algorithm as: (i) The crystallographic information files were analyzed to obtain the composition and symmetry information, including the elemental compositions, stoichiometry, the total atomic number of each element species, space group number, Hermann-Mauguin symbol, and Wyckoff symbol of each atom site; (ii) The structures were matched to find the duplicates. All compounds with the same composition (i.e., the same species and relative proportion) were compared using symmetry information. If the structures also had the same space group number, the same Hermann-Mauguin symbol, the same number of atoms at each Wyckoff position, and the same Wyckoff symbol of each corresponding atom site, then they were identified as duplicates. A unique structure was then selected from each duplicate group randomly as a representative for further calculations. Finally, we removed the crystallographic information file entries containing more than six elemental species.

**Crystal structure representation**. BCM: First, we chose the BCM as the LAE representation for the unsupervised learning models in step 1 clustering. The BCM encodes bond information for each atom with all its neighbouring atoms of the same species in a structure. It is an advanced version of the bond-orientational order parameter technique, first introduced by Steinhardt et al[52]. In the BCM, a bond vector $\vec{R}_{ij}$ between atoms *i* and *j* is defined when the interatomic distance between these two atoms is less than a cut-off distance. The bond orientation of $\vec{R}_{ij}$ is expressed by the spherical harmonics $Y_{lm}(\theta_{ij}, \phi_{ij})$, where $\theta_{ij}$ and $\phi_{ij}$ are polar angles. The average over all bonds formed by type A and B atoms can be derived by the equation

$$\bar{Q}_{lm}^{\delta_{AB}} = \frac{1}{N_{AB}} \sum_{i \in A, j \in B} Y_{lm}(\theta_{ij}, \phi_{ij}), \qquad (1)$$

where $\delta_{AB}$ and $N_{AB}$ denote the type and number of bonds, respectively. Only the even spherical harmonics ($l$) are used in Eq. (1) to ensure invariant bond information with respect to the direction of the bonds. Taking the rotationally invariant combinations into consideration, Eq. (1) is now defined as:

$$Q_l^{\delta_{AB}} = \sqrt{\frac{4\pi}{2l+1} \sum_{m=-l}^{l} \left|\bar{Q}_{lm}^{\delta_{AB}}\right|^2}, \qquad (2)$$

where each series of $Q_l^{\delta_{AB}}$ for $l$ = 0, 2, 4, 6, 8, and 10 can be used to represent a type of bond. The similarity of two structures can then be quantitatively represented by the Euclidean distance between their BCMs with the following formula:

$$D_{\mu\nu} = \sqrt{\frac{1}{N_{type}} \sum_{\delta_{AB}} \sum_l \left(Q_l^{\delta_{AB},\mu} - Q_l^{\delta_{AB},\nu}\right)^2}, \qquad (3)$$

where $\mu$ and $\nu$ represent two individual structures and $N_{type}$ is the number of bond types. Therefore, for any given structure, whether the number of atoms is the same or not, if the number of bond types (atomic types) is the same, then the similarity between them can be easily compared by the BCM Euclidean distance.

With the same bonding cut-off distance and the same local atomic environments, different lattice parameters may lead to different BCM results. To avoid this in step 1 clustering, we scaled all the structures to be with the nearest-neighbour distance 1.52 Å, which is the average value of all covalent radii defined by Cordero and his colleagues[53]. We use (1.52+Δ) Å as the bonding cut-off distance by encoding the nearest neighbour bonding information only (with Δ= 0.05). To satisfy the periodic boundary conditions of crystal structures, we use sufficiently large supercells to calculate the BCM matrixes. Only the bonds in the central unit of the supercell are calculated.

SOAP: The SOAP fingerprint[43] encodes a local atomic environment $\chi$ as an N×1 tensor (N is the number of atoms in $\chi$). Each element of the tensor is a sum of atom-centred Gaussian smeared atomic densities expended in a basis of spherical harmonics and a set of radial basis functions. Each atom in $\chi$ is centred on once. In the SOAP fingerprint, each LAE is first transformed into atomic density fields $\rho\langle\alpha\mathbf{r}|\chi\rangle$:

$$\rho\langle\alpha\mathbf{r}|\chi\rangle = \sum_{i \in \chi, \alpha} exp\left(-\frac{(X_i - r)^2}{2\sigma^2}\right), \qquad (4)$$

where $X_i$ are the Cartesian coordinates of the atoms of chemical species $\alpha$ within a radial cut-off $r_c$. $\mathbf{r}$ is the Cartesian coordinate of the central atom and $\sigma$ denotes the width of the Gaussian function.

After expansion with a set of orthonormal radial basis functions $g_n$ and spherical harmonic $Y_{lm}$, $\rho\langle\alpha\mathbf{r}|\chi\rangle$ can be expressed as:

$$\rho\langle\alpha\mathbf{r}|\chi\rangle = \sum_{nlm} c_{nlm}^{\alpha} g_n(r) Y_{lm}(\theta, \phi), \qquad (5)$$

where the coefficients are defined as:

$$c^{\alpha}_{nlm} = \iiint_{R^3} dV g_n(r) Y_{lm}(\theta, \phi) \rho \langle \alpha r | \chi \rangle. \tag{6}$$

To guarantee the rotationally invariant output of SOAP, the partial power spectrum vector $p(r)$ is implemented. The elements in $p(r)$ are defined as:

$$p(r)^{\alpha_1 \alpha_2}_{nn'l} = \pi \sqrt{\frac{8}{2l+1}} \sum_m c^{\alpha_1}_{nlm}(r)^* c^{\alpha_2}_{n'lm}(r). \tag{7}$$

Then, the SOAP kernel between two atomic environments can be defined as a normalized polynomial kernel of the partial power spectrum:

$$K(p, p') = \left( \frac{p \cdot p'}{\sqrt{(p \cdot p)(p' \cdot p')}} \right)^{\xi}. \tag{8}$$

The SOAP kernel distance $d(\chi, \chi')$ can ultimately be defined as:

$$d(\chi, \chi') = \sqrt{2(1 - K(p, p'))}. \tag{9}$$

Combining all the local atomic environment descriptors, we can obtain a global measure of the similarity between two structures (A and B) with the same number of atoms $N$ and the same number of atomic types. We compute an environment covariance matrix containing all the possible pairings of environments:

$$C_{ij}(A, B) = K(\chi^A_i, \chi^B_j). \tag{10}$$

Now, the key to measuring the similarity between structures A and B is finding the best match between pairs of environments in the two configurations. Therefore, we use a global kernel called the "regularized entropy match kernel" (REMatch kernel) defined by Sandip et al.[54] to calculate the global SOAP kernel between structures A and B. The regularized entropy match kernel is defined as:

$$K(A, B) = Tr \mathbf{P}^{\alpha} \mathbf{C}(A, B), \tag{11}$$

$$\mathbf{P}^{\alpha} = \underset{P \in \mho(N,N)}{argmax} \sum_{ij} P_{ij}(1 - C_{ij} + \alpha \ln P_{ij}), \tag{12}$$

where $\mho(N, N)$ is the set of $N \times N$ doubly stochastic matrixes and $\sum_i P_{ij} = \sum_j P_{ij} = \frac{1}{N}$.

Finally, the global SOAP kernel distance (similarity) between structures A and B can be defined as:

$$D(A, B) = \sqrt{2(1 - K(A, B))}. \tag{13}$$

The REMatch kernel can also be computed easily for a rectangular matrix, which constitutes an additional advantage of formulating the environment matching problem in terms of regularized transport optimization. Here, by evaluating the least common multiple $N$ of $N_A$ and $N_B$ and replicating the environment similarity matrix to form a square matrix, using the REMatch kernel can easily and more efficiently compare the similarity between two structures with different numbers of atoms.

In step 2 clustering, we used the global SOAP kernel distance matrix to measure the similarities between different structures. The same as in step 1 clustering, for eliminating the effect of the scale of the lattice parameters, we also scaled the lattice parameters. After scaling, the nearest-neighbour distance of each structure is set to 1.52 Å. To compromising the computational cost and the computational accuracy, we use an empirical value of 4.56 Å (three times of 1.52 Å) as the cut-off radius. Another unique work can be done to perform the grid search testing according one's own discretion. To eliminate the possible effect of the different element species on the SOAP results, we replaced the element species list for each structure with the same list of element species. Specifically, for unary compounds, we used "C"; for binary, we used "C, Ge"; for ternary, we used "C, Ge, O"; for quaternary, we used "C, Cl, Ge, O"; for quinary, we used "C, Cl, Ge, K, O"; and for senary, we used "B, C, Cl, Ge, K, O". The types of element species have no effect on the SOAP results as long as all the structures in the same group have the same element species list. Considering the periodic boundary conditions of crystal structures, we also used sufficiently large supercells to calculate the SOAP fingerprints. We used the *SOAP* module and *REMatchKernel* function from the *DScribe* package[45] to calculate the SOAP fingerprints and SOAP kernel distance matrix. For both BCM Euclidean distance calculations and SOAP kernel distance calculations, we considered all the permutating orders of the replacing atoms for each structure, and chose the shortest distance between the two compared structures as the distance between these two structures. For materials with Li-Co-O species, all the six possible substitution cases including C-Ge-O, C-O-Ge, Ge-C-O, Ge-O-C, O-C-Ge, O-Ge-C were considered.

**Unsupervised learning**. We used the *linkage*, *kmeans*, and *dendrogram* functions from the open-source *SciPy* package[55] to perform agglomerative hierarchical clustering and *k*-means clustering. Agglomerative hierarchical clustering begins with an initial set of singleton clusters, including all the objects. Each sample belongs to its own cluster. Then, the pair of clusters with minimum dissimilarity is agglomerated as a new cluster. The agglomeration step is repeated progressively and greedily according to the similarity metric until the number of clusters equal to the clustering threshold value or the clustering distance reaches the cut-off value. The set of clusters from agglomerative hierarchical clustering process forms a bottom-up hierarchical clustering tree diagram (dendrogram). In step 1 clustering, the Euclidean distance (L2) between two BCM matrixes is modelled as a similarity metric. In step 2 clustering, the SOAP global kernel distance between two SOAP matrixes is modelled as a similarity metric. In both step 1 and step 2 clustering, the unweighted average linkage is used to measure the cluster dissimilarity[47,48]. To the beginning, we have also considered the "single", "complete", "centroid", "ward" (Ward's) linkage methods. We have examined the effect of these different linkage methods on the selected groups of $(A_1B_1C_{10})_2$. The results show that the linkage method used in the clustering algorithm does little effect for our clustering results. In this work, we are aimed at constructing an average version of structure prototype database. Therefore, we chose "average" linkage methods for clustering. (see "Supporting Information" to get more details).

**DFT calculation**. We used our in-house developed code, the Jilin artificial intelligence-aided material design integrated package (*JAMIP*)[20,56] to perform all the DFT-based high-throughput calculation tasks, and analyze the calculated results. The VASP package were used to perform all the DFT calculations. The plane-wave pseudopotential methods implemented in VASP were adopted. The electron-ion interaction was described by the projected augmented pseudopotential method. We used the Perdew-Burke-Ernzerhof-generalized gradient approximation[57,58] as the exchange-correlation functional to relax all the crystal structures and calculate the band gaps of $MgCl_2$ structures. For the prototype structures stored in the LAE-ICSPD, a kinetic energy cut-off of $1.3 \times E_{max} eV$ was used for plane wave expansion, where $E_{max}$ is the maximum cut-off energy for the plane wave basis of all species within a structure. For the $MgCl_2$ system, a kinetic energy cut-off of 341 eV was used. For prototype structures relaxation, we limited the k-point meshes with a grid spacing of $2\pi \times 0.03 Å^{-1}$ for Brillouin zone integration. The equilibrium structural parameters (including the lattice parameters and internal coordinates) were optimized via total energy minimization with a convergence criterion for the residual force on each atom of less than $0.01 eV/Å$ (for $MgCl_2$, we used $0.001 eV/Å$).

**Python scripts**. We integrated all the Python scripts to form SPGI. The LAE-ICSPD was constructed with the help of SPGI. We successively used the commands or Python scripts below to identify the structure prototypes from the initial structure dataset:

❖ FILTER
- The command "python3 rmpocp.py" was used to remove the structures with partial occupancy.
- The command "python3 rmmissatom.py" was used to remove the structures with missing atoms (e.g., H).
- The command "python3 rmutradis.py" was used to remove structures with ultrasmall atomic distances.
- The command "python3 rmdup-symmery.py" was used to remove the structures with the same formula and symmetry.
- The command "python3 attnum6.py" was used to remove structures with more than six species.

❖ PTP1-ATMTYPE
- The command "python3 ptp1-atmtype.py" was used to classify the structures according to the number of atomic types (species).

❖ PTP2-SIMSTOICH
- The command "python3 ptp2-simstoichA/B/C/D/E/F.py" was used to classify the structures in each PTP1-ATMTYPE group according to their simplest stoichiometry. The corresponding classification results were placed into groups A, AB, ABC, ABCD, ABCDE, and ABCDEF.

❖ PTP3-UNITNUM
- The command "python3 ptp3-unitnum.py" was used to classify the structures in each PTP2-SIMSTOICH group according to the total number of atoms (i.e. number of formula units).

❖ PTP4-BCM_HC

- The command "python3 get_bcm_Eucd.py" was used to calculate the BCM Euclidean distance matrixes of each group in PTP3-UNITNUM.
- The command "python3 run_hc_bcm.py" was used to run the agglomerative hierarchical clustering processes in step 1 to hierarchically cluster the structures in each PTP3-UNITNUM group based on their BCM Euclidean distance matrixes.

❖ PTP5-SOAP_HC
- The command "python3 get_soap_Knld.py" was used to calculate the SOAP kernel distance matrixes of each clustering group in PTP4-BCM_HC.
- The command "python3 runhc_soap.py" was used to run the agglomerative hierarchical clustering processes in step 2 to hierarchically cluster the structures in each PTP4-BCM_HC group based on their SOAP kernel distance matrixes.

❖ LAE-ICSPD
- The command "python3 getpt.py" was used to call "km_cluster.py" to obtain the final structure prototypes with the help of $k$-means clustering algorithm.
- The command "python3 lcspd_nm.py" was used to systemically allocate the LAE-ICSPD collecting code and list the other structure information for each structure prototype stored in LAE-ICSPD.

## 3. Results

**Overviewing workflow of constructing the LAE-ICSPD.** By utilizing the power of unsupervised learning for large-scale structure clustering, we followed three steps to construct the LAE-ICSPD (Figure 1). First, by extracting all the experimentally known inorganic crystal structures from the ICSD, we obtained 193,466 inorganic compounds as the initial dataset. Next, by filtering the initial dataset with five sequential filtering criteria (i.e., partial occupancy, missing atoms, ultrasmall atomic distance, same formula, same symmetry, and containing more than six elements), we identified 55,430 effective inorganic crystal structures. Then, we classified these effective structures by their stoichiometric information. Finally, by using unsupervised learning frameworks, we selected 15,613 inorganic crystal structures as the structure prototypes. After being geometrically relaxed by DFT-based high-throughput calculations, the relaxed prototype structures were numbered systematically and stored in the LAE-ICSPD.

**Pre-processing of the effective experimentally known inorganic crystal structures**. With the aim of obtaining high-quality experimental data and accelerating the unsupervised learning models, we filtered and pre-processed the initial structure dataset as shown in Figure 1. The filtering process starts from an initial dataset containing 193,466 entries. All entries are unit cells in the form of crystallographic information file and are known experimentally. First, we excluded structures with partial occupancies in the unit cell. This step reduced the number of entries to 102,832. Then, we discarded the structures with missing atoms (e.g., H) and the structures with ultrasmall atomic distances reducing the number of entries to 93,450 and 93,196. Next, we eliminated duplicate structures with the same structural formula and symmetry. This step reduced the number of entries to 55,753. Finally, we retrieved the structures containing a maximum of six different elemental species. The filtering protocol reduced the initial set of 193,466 structures to 55,430 effective structures (see "Methods"). Here, we have made every attempt to remove inconsistent structures from the database. Automated filtering is not sufficient as it renders different types of errors and partial structures in the ICSD. Therefore, the numbers presented here should be considered as an estimate.

We then classified the effective structures using their stoichiometric information. These structures were first divided into unary compounds labeled as A, binary compounds as AB, ternary compounds as ABC, quaternary compounds as ABCD, quinary compounds as ABCDE, and senary compounds as ABCDEF according to the number of trace element species. Next, the structures were classified into 4653 sub-groups based on their simplest chemical stoichiometric ratios. Moreover, the smaller size of the structure similarity matrixes can significantly reduce their computational cost. In addition, to ensure the LAEs' comparative fairness and to generate more hierarchical clustering analysis groups, the structures in each simplest stoichiometric ratio sub-group were further classified into many groups based on the total number of atoms (actually, the highest common factor of the stoichiometry, also called the 'formula unit') in their unit cells. We initially ran the unsupervised learning frameworks relying on each of this kind of sub-group to identify at least one structure prototype from each group.

**Representation of LAEs based on the state-of-the-art structure fingerprints**. To run unsupervised learning models with high computational efficiency and make the models learn to accurately measure the similarity between different structures, a set of quantitative and effective representations of the complex material structures are required as input. For the appropriate representation of LAEs is a crucial component of the algorithms used in modern computational materials science[43]. Both BCM and SOAP are structure fingerprints designed to describe the LAEs of material structures. Therefore, we chose BCM and SOAP fingerprints to represent the LAEs of each structure.

BCM starts by considering the bond order information (i.e., the bond angle) of all the bonds in a structure. This is a variant and improved version of the bond-orientational order parameter technique. The bond-orientational order parameter is usually employed to characterize the local environment of a central atom[59,60]. All the neighbouring atoms of the same species are considered forming bonds with it. BCM generalizes the bond-orientational order parameter to multi-component coupling; i.e., the atomic type or bond type is encoded. In this work, we generalize the BCM to crystal structures and calculate the BCM Euclidean distance matrix to measure the similarity between different structures (see "Methods").

As the BCM fingerprint does not contain bond length information, structures with the same bond angles but different bond lengths may not be distinguished accurately. Herein, we employ the SOAP fingerprint as the second characterization to handle this problem for SOAP encoding of both the bond angle and bond length information. In addition, SOAP is sensitive to the atomic type (bond type). It is a continuous and differentiable distance metric that describes the local atomic environment of an atom as a sum of atom-centred Gaussian functions. Then the overlap of two local atomic environments (densities) integrated over all three-dimensional rotations is defined as the SOAP kernel. We calculate the SOAP kernel distance matrix to further measure the similarity between different crystal structures (see "Methods").

**Unsupervised learning of LAEs through hierarchical clustering algorithm.** As we did not know how many clusters we should group in our structure dataset, we performed hierarchical clustering, which is a common unsupervised learning technique that does not assume the cluster number of the dataset in advance. As Figure 1 shows, in step 1 clustering, we first ran bottom-up agglomerative hierarchical clustering processes based on each BCM Euclidean distance matrix. The grouping results revealed a fine quality of clustering, as the structures shared similar characteristics within the same groups. As Figure 2 shows, the six structures with a stoichiometric ratio of 1:1:10 ($A_1B_1C_{10}$) were clustered into two different BCM-based clusters (BC1 and BC2) in step 1 clustering. Here, the structures in the same cluster shared the same or very similar geometry and LAEs. Interestingly, in some groups, the crystal structures were classified automatically based on their space group (the upper half of Figure 2). The two structures with formula $(CuClF_{10})_2$ belonging to space group *$P2_1/c$* or *$P2_1/a$* were clustered into BC1, while the remaining four structures with formulas $(UPCl_{10})_2$, $(NbSbF_{10})_2$, $(NbPCl_{10})_2$, and $(TaPCl_{10})_2$, sharing the same space group (*P-1*), were clustered into BC2. We note that the fundamental purpose of this clustering step is to split the clustering structure space into smaller volumes to accelerate the next step of clustering process. Therefore, in step 1 clustering, we used the computationally efficient descriptor BCM. We also set a primary cut-off point by finding it within the upper half of the clustering dendrogram to obtain as few clusters as possible. This can prevent structures belonging to the same structure prototype being divided into different clusters in this step.

In step 2 clustering, to further measure the structural similarities in BCM-based clusters, we ran the bottom-up agglomerative hierarchical clustering processes using the SOAP kernel distance matrixes as input. The SOAP descriptor is more accurate than the BCM descriptor but incurs much greater computational cost than the BCM descriptor. In practice, calculating the SOAP kernel distance matrix is a very time-consuming task compared to calculating BCM Euclidean distance matrix. In step 2 clustering, we directly relied on each BCM-based cluster's SOAP kernel distance matrix to run the hierarchical clustering processes to accelerate the clustering analysis. After being clustered in step 2, the two structures in BC1 were still grouped into the same SOAP-based cluster (BC1/SC1) due to the much shorter SOAP kernel distance between these two structures. The four structures in BC2 were divided into two different groups (BC2/SC1 and BC2/SC2). The structure $(NbSbF_{10})_2$ in BC2/SC1 has much different lattice geometries and LAEs compared to the other three structures in BC2/SC2. The structures in the same SOAP-based cluster shared quite similar lattice geometries and LAEs, while the structures in different SOAP-based clusters were much different (lower panel in Figure 2). In step 2 clustering, we set a refined cut-off point by finding it over all the clustering dendrogram to get a suitable clustering threshold value in each clustering group. Finally, we used the k-means clustering algorithm to identify one structure prototype from each cluster in terms of the SOAP structure characterization criterion. In this work, the k-means clustering algorithm was only used to find the central point of each SOAP-based cluster. Then we selected the structure closest to the central point as the structure prototype. Considering the computational efficiency, we ran it based on the BCM Euclidean distance matrix. There are no structures that do not belong to any structure prototype, and each singleton cluster represents a new prototype.

After clustering analysis by our unsupervised learning frameworks, we found that not only some structures with the same space group were divided into different structure prototypes (due to their different LAEs), but also some structures with different space groups were clustered into the same structure prototype due to their identical or nearly similar LAEs. Moreover, Finding a suitable clustering cut-off point in a hierarchical clustering model is the key to obtain as many clusters with high uniqueness as possible, which can guarantee the high uniqueness and variety of the structure prototypes. There exist some methods that determine the clustering cut-off point (also called clustering threshold value) in a dataset, including the elbow method and the inconsistency method. Due to the obvious variability and diversity of our dataset, each of them cannot be qualified for all cases. Here, we defined a new method called the "compromised maximum growth rate" to choose the clustering threshold value automatically in all cases (see Figure S2). In this method, the clustering threshold value is co-determined by the statistical properties of the total dataset and the current sub-dataset. (More details about this method can be found in the "Supporting Information".)

We integrated all the above three function modules into SPGI. It is different from other similar packages which use the space group information of crystal structures as the predominant criterion for classification[41,42,61–63]. SPGI directly identifies the structure prototypes by learning the local atomic environments of all the crystal structures. SPGI first discards the ineffective structures, then classifies the effective structures by their stoichiometric information, and hierarchically clusters the structures based on the unsupervised learning of LAEs. The unsupervised learning framework in SPGI is flexible. It allows variation of the local atomic environment representation and the clustering cut-off point search range, which may lead to different structure prototype results to meet one's own discretion. SPGI can be easily extended to other local atomic environment representation matrixes such as many-body tensor representation[64], atom-centred symmetry functions[65], Coulomb matrixes[66], etc. (see "Methods" to get more details about SPGI).

**Constructing the LAE-ICSPD.** With the help of SPGI, we first retrieved all the effective structures, then classified them into six groups according to the number of atomic types. These six groups were stored in AB… class. Then, classifying all the structures in each AB… group by the simplest stoichiometric ratio, we obtained many sub-groups stored in the $A_xB_y$… class. Next, we classified all the structures in each $A_xB_y$… group by the formula unit and obtained more sub-groups stored in the $(A_xB_y…)_U$ class. Finally, being clustered by our unsupervised learning frameworks, we identified 15,613 structure prototype groups stored in the $(A_xB_y…)_U$-ID class. The total number of structures or groups in each class is listed in Table 1. Refined by the *k*-means clustering algorithm, we ultimately obtained 15,613 structure prototypes from the $(A_xB_y…)_U$-ID class. For the structure prototypes containing 100 or fewer atoms, we ran DFT-based high-throughput calculations (see

"METHODS") to optimize the structures through total energy minimization. For other complex structures, the structure data identified through experiments were taken. We generated a unique collection code for each structure prototype. Each unified collection code number consists of the simplest stoichiometry, the total number of atoms in the unit cell and the unique collection number in the LAE-ICSPD. For example, the 3846th structure prototype in the LAE-ICSPD is a phase with the chemical formula "$Pb_8Ti_8O_{24}$". It belongs to the $ABC/A_1B_1C_3/(A_1B_1C_3)_8$ group. Here, the LAE-ICSPD collection code is "$A_1B_1C_3\_40\_3846$". The collection code and other structural information (space group, etc.) of each structure prototype in the LAE-ICSPD are collected in a list.

Figure 3 shows the distribution of the data collection process for the LAE-ICSPD. The number of total structure prototypes against the total number of experimentally known structures in each atomic type group is depicted in Figure 3a. The number of structure prototypes per 100 experimentally known crystal structures in each atomic type group is approximately 18 (A), 13 (AB), 17 (ABC), 40 (ABCD), 70 (ABCDE), and 86 (ABCDEF). This indicates that more complex crystalline structures may possess much different LAEs. Moreover, the sum of the structure prototypes in the ternary (ABC) and quaternary (ABCD) groups accounts for more than 60% of the total number of structure prototypes in the LAE-ICSPD database.

By comparison, we constructed other inorganic crystal structure prototype database based on the space group of the crystal structures (named as SG-ICSPD). Figure 3b shows the number of structure prototypes in LAE-ICSPD against the number of structure prototypes in SG-ICSPD of each atomic type group. The total numbers of structure prototypes in both of the cases are almost the same. For the former, it is 15,613. For the latter, it is 16,058. This validates (to some extent) the rationality of the structure prototype extraction results of our unsupervised learning frameworks. In the groups within three atomic types, the number of space group-based prototypes is greater than the number of LAE-based prototypes. When starting from the quaternary group, the situation is the opposite. This also indicates that complex crystalline structures may possess much different LAEs.

**Table 1.** The number of total structures or groups in each class.

| AB… | Structure | $A_xB_y…$ | $(A_xB_y…)_U$ | $(A_xB_y…)_U$-ID |
|---|---|---|---|---|
| **A** | 434 | 1 | 43 | 77 |
| **AB** | 10,243 | 215 | 589 | 1,352 |
| **ABC** | 25,664 | 811 | 1,704 | 4,452 |
| **ABCD** | 12,996 | 1,303 | 2,323 | 5,256 |
| **ABCDE** | 4,761 | 1,505 | 2,110 | 3,333 |
| **ABCDEF** | 1,332 | 818 | 937 | 1,143 |
| **TOTAL** | 55,430 | 4,653 | 7,706 | 15,613 |

For computational material screening and design studies, compared to the substitution-based enumeration approaches, taking the representative and effective structural prototypes to generate candidate material structures will dramatically reduce the duplicate or ineffective structures to be calculated, herein lower the computational cost. Taking the AB-based stoichiometry as the instance, without any structural prototypes database, for the substitution-based enumeration step one need to calculate totally 10,243 structures in the ICSD for a complete high-throughput calculations. By using our developed LAE-ICSPD, only 1,352 effective structures need to be considered.

**Benchmarking the constructed LAE-ICSPD by using the $A_1B_2$-based stoichiometry.** We benchmarked the constructed LAE-ICSPD using the $A_1B_2$-based stoichiometry, *i.e.*, taking $MgCl_2$ as the compound to be investigated. We first retrieved all the 2,387 experimentally known inorganic crystal structures from the ICSD with no more than 30 atoms in $A_1B_2$ group as the possibly existing candidate structures. Then we substituted the atoms at A-site by Mg cation, the B-site atoms by Cl anion to generate $MgCl_2$ structures. Next, we relaxed the crystalline lattice parameters and atomic positions of all these $MgCl_2$ structures using DFT-based high-throughput calculations with our in-house developed code, the Jilin artificial intelligence-aided material design integrated package (*JAMIP*)[20,56]. After removing the structures not converged in the total energy minimization, we remained 2,087 effective structures. We calculated their cohesive energy (per formula) and band gaps as the function of the quantity to evaluate volume density, packing factor.

The calculated results are illustrated in Figure 4. The 2,087 optimized $MgCl_2$ structures (gray dots in Figure 4) is labelled as "ALL-EXP". In case of our developed LAE-ICSPD adopted, only 69 structures (blue cross) are needed to be calculated. By comparison, we have also adopted the SG-ICSPD, from which 174 structures (green circle) need to be calculated. We found that although the data of both LAE-ICSPD and SG-ICSPD do not entirely cover all the data in the ALL-EXP, both of prototype databases lead us to identify the experimentally known and thermodynamically stable structure of $MgCl_2$[67] (marked by the black star). With the different strategies to produce the structure prototypes, the candidate structures to be calculated in the high-throughput calculations show clear difference. With fewer number of the structure prototypes, our developed LAE-ICSPD efficiently cover the region with relatively low cohesive energy (Figure 4a). Interestingly, we found 11 $MgCl_2$ structures showing quite different packing factors, but having comparable energy to the experimentally reported structure of $MgCl_2$. We can further adjust the clustering cut-off value in our hierarchical clustering models to produce flexibly the ICSPD with more LAE-based structure prototypes. By reducing the clustering cut-off value (*i.e.*, scaling down the average value of the all maximum/average/minimum clustering distances of each group by one hundredth times), we can obtain a $A_1B_2$-based ICSPD with 883 structure prototypes (red cross, labeled as "LAE′-ICSPD"). As expected, the data of LAE′-ICSPD covered more representative structures in the ALL-EXP. This flexibility of our developed LAE-ICSPD, *i.e.*, providing possibility of generating structure prototypes with the tunable clustering criteria, is expected to facilitate the high-throughput calculations or computational materials by design studies with balancing computational afford and targeted goal. We have performed further statistical analysis on the correlation between the energy difference and the structural difference (Figure S7). By using our developed LAE-ICSPD and LAE′-ICSPD, one can efficiently sampling the corners of the potential surface with continuously varying structural difference. This is particularly evident in the low energy region where the thermodynamically stable and metastable structures exist. This dramatically reduces the number of DFT calculations and accelerates the screening process. For computational crystal structure prediction, sampling representative structures and meanwhile having diversity in the structures generation steps will significantly accelerate the convergence to the global minimization solution and enhances the structure search efficiency. The above results imply that there exist the structures with the same space group and different LAE and the ones with different space groups and the same LAE as discussed before. Therefore, for creating new materials in the computational materials by design study, one need to cautiously use the structure prototype database exhausting all the potential LAEs to perform high-throughput DFT calculations for avoiding omitting possible candidate structures.

We note that in the above benchmark calculations by simply taking $MgCl_2$, adoption of the LAE-ICSPD allows us to identify potential metastable $MgCl_2$ structures with reasonably low cohesive energy and significant change in band gap values in rather wide range. It is expected to apply the LAE-ICSPD to find the new metastable materials with target stoichiometry with emerged useful properties that can be experimentally realized by the non-equilibrium growth method or under the extreme condition (e.g., intensely strained, high pressure, *etc.*).

## 4. Discussion and conclusion

In this work, we have constructed an inorganic crystal structure prototype database LAE-ICSPD based on unsupervised learning of local atomic environments of all effective experimentally known inorganic crystal structures. Particularly, we have adopted hierarchical clustering approach onto these structures to identify structure prototypes. Different from previous space group based works, the structure prototypes in LAE-ICSPD have been identified by directly and intelligently comparing the LAE-based similarities between different structures with the help of unsupervised machine learning algorithms, without classifying them according to their space groups. Herein, the structure prototypes in LAE-ICSPD owns higher uniqueness. Under our LAE-based unsupervised learning frameworks, not only are structures with very different LAEs and the same space group automatically divided into different structure prototypes, but also, structures with identical LAEs and different space groups are automatically clustered into the same structure prototype. It should be pointed out that another key motivation of identifying the unique structure prototypes with the help of the hierarchical clustering is to understand the chemical and structural diversity of the given database. Essentially, the unique structure prototypes are picked out by identifying the types of preferred structures by nature. Understanding such diversity will be beneficial for designing/discovering from a global angle of view advanced materials with emergent new functionality. We have also benchmarked the LAE-ICSPD by using the $A_1B_2$-based stoichiometry to generate virtual $MgCl_2$ structures and calculate their cohesive energies, band gaps, and packing factors. The local atomic environments of structures are represented by the state-of-the-art structure fingerprints including the improved bond-orientational order parameters and the smooth overlap of atomic positions. The similarities between different structures are measured by the BCM Euclidean distance matrixes and SOAP kernel distance matrixes. Moreover, we have developed a Structure Prototype Generator Infrastructure (SPGI) package for structure prototypes generation. Both the LAE-ICSPD and SPGI toolkit are useful for searching the material properties of inorganic materials in a global way and accelerating new materials discovery processes in data-driven modes, such as high-throughput computational materials design frameworks, data-driven machine learning based models, and computational crystal structure prediction models.

The LAE-ICSPD was constructed by first classifying the structures according their stoichiometries. Actually, using the local atomic environment descriptors based unsupervised learning framework in SPGI to generate structure prototypes allows one to directly compare the similarities between different structures without classifying them by their stoichiometries. However, we found that classifying the structures by their stoichiometries is necessary in most materials research fields[68,69]. This step can further split the clustering structure space into smaller volumes to accelerate the next step of clustering process. We herein performed this step. On the other hand, the unsupervised learning framework in SPGI is flexible. It allows variation of the local atomic environment representation and the clustering cut-off point search range, which may lead to different structure prototype results. It is still an open question when the hierarchical clustering process ends or which structure representation matrix should be applied to measure the similarities between different structures. Generally, SPGI allows one to flexibly operate the structure prototype generating results according to their own requirements. Furthermore, SPGI can be easily extended to other local atomic environment representation matrixes. As the results may vary with the cut-off parameters, one should do the grid search testing. But the grid search testing about the cut-off values of bonding radius, different structure representation matrixes, different calculating methods of distance may make no sense without considering one's own requirement. And this kind of grid search testing for the big data of initial structural dataset is pretty complex. We herein did not do the grid search testing in this work. It may be done by another unique work according one's own discretion. In this work, we only gave out a suitable set of parameters, based on which, constructed a basic version of inorganic crystal structure prototype database. With the help of SPGI, one can easily expend this database or construct another version of sub-databases based on different parameters to satisfy one's own needs.

We have considered employing our methodology also to explore structure prototypes of 2D materials. By using suitable 2D structure fingerprints to describe the 2D structures and measure the similarity between different 2D structures, the unsupervised learning methodologies in SPGI can also be straightforward applied to explore the structure prototypes of 2D materials from existing 2D materials databases, such as the Computational 2D Materials Database[70] (C2DB), the Virtual 2D Materials Database[23] (V2DB), 2D Materials Encyclopedia[71] (2DMatPedia) and Materials Cloud Archive[68], even from 3D materials database. When exploring the structure prototypes of 2D materials from 3D materials, one should use structure dimensionality identification methods, such as topology-scaling algorithm[72], rank determination algorithm[68], scoring parameter[73] et al. to pick out the exfoliated 2D materials from 3D materials databases firstly. However, there are some challenges for employing our methodology to explore structure prototypes of 2D materials. On the one hand, the experimentally known 2D compounds are rare, which can't provide enough initial 2D material dataset with high data quality for generating more 2D material prototypes. The 2D materials databases mentioned above are all quantum computational simulation based databases. On the other hand, using suitable 2D structure fingerprints to describe the 2D structures and measure the similarity between different 2D structures is essential to use our methodology to explore structure porotypes of 2D materials. Essentially, the dimensionality of one structure is determined by the topological connectivity of atoms. Herein, the structure fingerprints used to measure the similarity between different 2D structures for generating 2D structure prototypes must encode the topological connectivity information of atoms. Quotient graph may be a good choice[74].

## Acknowledgments

This work was supported by the National Natural Science Foundation of China (Grants No. 62125402, 92061113, 12004131, 62004080) and the Interdisciplinary Research Grant for PhDs of Jilin University (Grants No. 101832020DJX043). Calculations were performed in part at the high-performance computing center of Jilin University, China.

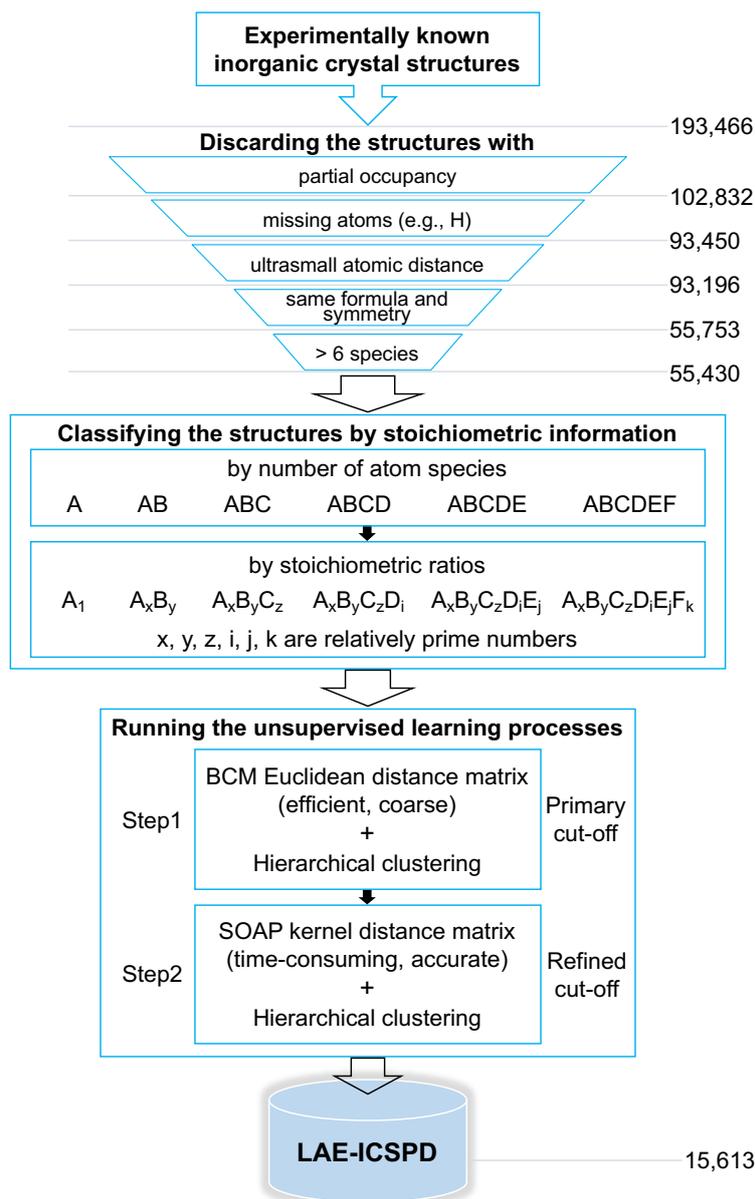

**Figure 1.** The workflow of constructing LAE-ICSPD. The effective structures are first filtered out from all the experimentally known inorganic crystal structures. Then these effective structures are classified by the stoichiometric information. Next, the unsupervised hierarchical clustering strategy is employed to get the final structure prototype clusters. The primary cut-off means finding the clustering cut-off point within the upper half part of the clustering dendrogram to get as few clusters as possible, while the refined cut-off means finding the cut-off point over all the clustering dendrogram to get a suitable number of clusters.

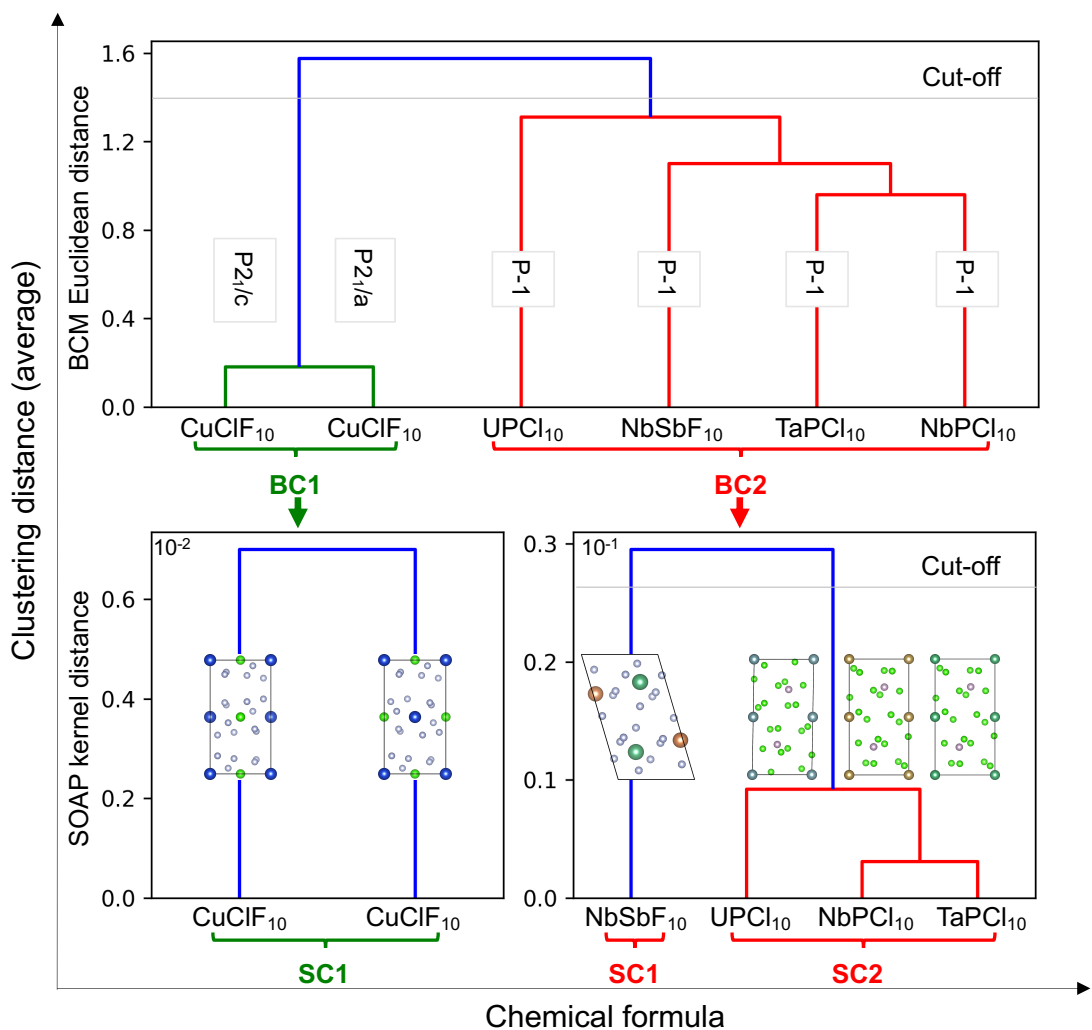

**Figure 2.** The hierarchical clustering dendrogram of $(A_1B_1C_{10})_2$ group. All the six structures are first grouped into two clusters (BC1 and BC2), based on the bond characterization matrix (BCM) Euclidean distances. Based on the smooth overlap of atomic positions (SOAP) kernel distances, the two structures in BC1 are grouped into cluster BC1/SC1, the four structures in BC2 are grouped in cluster BC2/SC1 and BC2/SC2. BC represents the clusters generated based on BCM whereas SC shows the clusters generated based on SOAP. In both of steps, the "average" linkage method is adopted to compute the distance between two clusters.

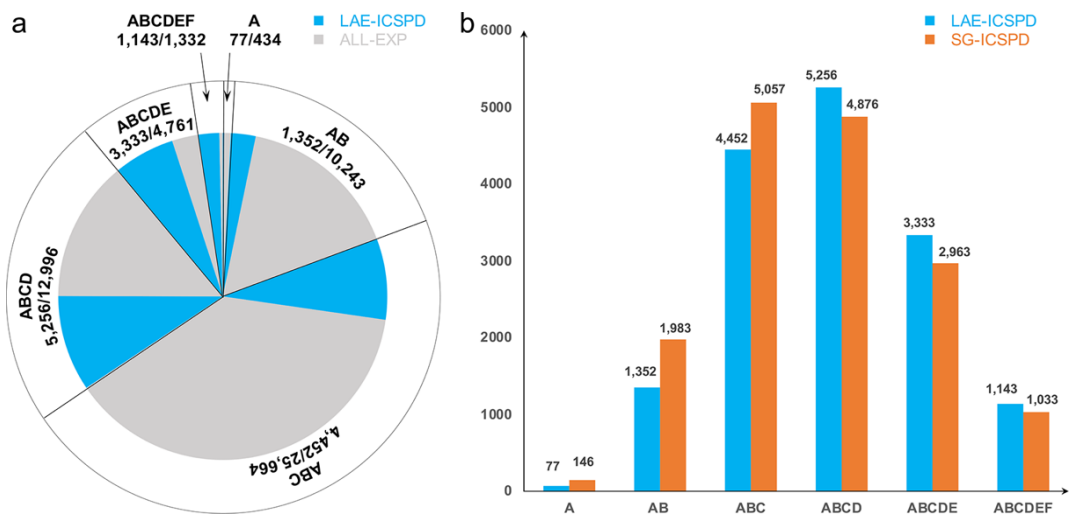

**Figure 3.** The statistical diagram showing number of structures in each atomic type group. ALL-EXP: The all experimentally known structures; LAE-ICSPD: The prototype structures stored in our local atomic environment (LAE) based inorganic structure prototype database (ICSPD); SG-ICSPD: The prototype structures stored in the space group (SG) based structure prototype database. The left panel (Figure a) shows the sector diagram of the total structure number of each atomic type group in LAE-ICSD (blue) or ALL-EXP (gray). The right panel (Figure b) shows the bar diagram of the total structure prototype number of each atomic type group in LAE-ICSD (blue) or SG-ICSPD (orange).

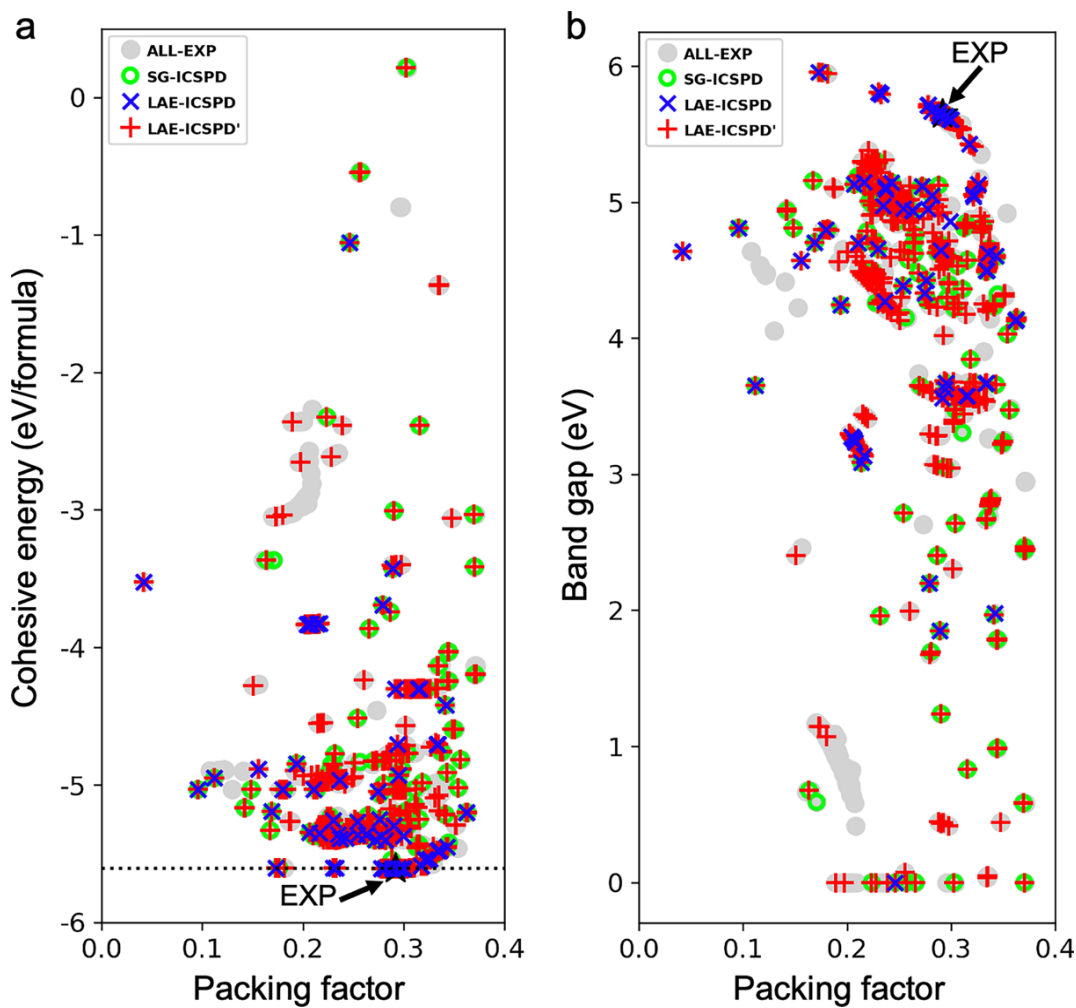

**Figure 4.** The scatter diagram analysis of MgCl$_2$ structures with respect to cohesive energy (Figure a) or bandgap (Figure b) against packing factor. ALL-EXP: The all effective MgCl$_2$ structures generated by substituting the elements in all experimentally known A$_1$B$_2$ structures (within 30 atoms in the unit cell); LAE-ICSPD: The MgCl$_2$ structures generated by decorating the A$_1$B$_2$ structure prototypes stored in our local atomic environment (LAE) based inorganic structure prototype database (ICSPD); LAE-ICSPD′: The other set of MgCl$_2$ structures generated by decorating the LAE based structure prototypes. This set of structure prototypes are generated by scaling down the average value of the all maximum/average/minimum clustering distances of each group by one hundredth times to get more structure prototypes; SG-ICSPD: The MgCl$_2$ structures generated based on structure prototypes stored in the space group (SG) based structure prototype database. The number of total structure prototypes in ALL-EXP, SG-ICSPD, LAE-ICSPD, LAE-ICSPD′, is 2,087, 174, 69, 883 respectively. EXP means the experimentally known MgCl$_2$ structures (marked by black stars). In Figure a, the black horizontal dot line is at the cohesive energy level of the experimentally known MgCl$_2$ structure.

# Supporting Information for "Inorganic Crystal Structure Prototype Database based on Unsupervised Learning of Local Atomic Environments"

Shulin Luo[1], Bangyu Xing[1], Muhammad Faizan[1], Jiahao Xie[1], Kun Zhou[1], Ruoting Zhao[1], Tianshu Li[1], Xinjiang Wang[2], Yuhao Fu[2,3], Xin He[1], Jian Lv[2,3], Lijun Zhang[1,3,*]

[1]State Key Laboratory of Integrated Optoelectronics, Key Laboratory of Automobile Materials of MOE, School of Materials Science and Engineering, and Jilin Provincial International Cooperation Key Laboratory of High-Efficiency Clean Energy Materials, Jilin University, Changchun 130012, China

[2]State Key Laboratory of Superhard Materials, College of Physics, Jilin University, Changchun 130012, China

[3]International Center of Computational Method and Software, Jilin University, Changchun 130012, China

*Corresponding author: Lijun Zhang, lijun_zhang@jlu.edu.cn

**This supporting information presents the following contents:**

- Example for structures with different space group owning the same LAEs
- The "compromised maximum growth rate" method
- The clustering results of $(A_1B_1C_{10})_2$ group based on different linkage methods
- The correlation between the energy difference and the structural distance for $MgCl_2$ structures
- Supporting references

- **Example for structures with different space group owning the same LAEs**

Figure S1 shows three structures with a stoichiometry of 1:1:8, including two $(BNF_8)_4$ phases that are synthesized at different temperatures and one $(FePCl_8)_4$ phase[1]. The $(BNF_8)_4$ phase, synthesized at 293 K belongs to the *P-42$_1$m* space group[2]. The other phase, synthesized at 153 K, belongs to the *Pbcm* space group[3]. The later has the same space group as the $(FePCl_8)_4$ phase. Although the two $(BNF_8)_4$ phases belong to different space groups, they have very similar or even identical LAEs, as displayed in Fig. 1a. In spite of the same space group, the $(FePCl_8)_4$ phase has quite different LAEs compared to the *Pbcm* $(BNF_8)_4$ phase (Fig. 1b). This indicates that the atomic type-sensitive LAEs encode both atomic site and atomic type configuration information, hence can effectively describe the global atomic environment, which is the essential criterion for structure prototype classification.

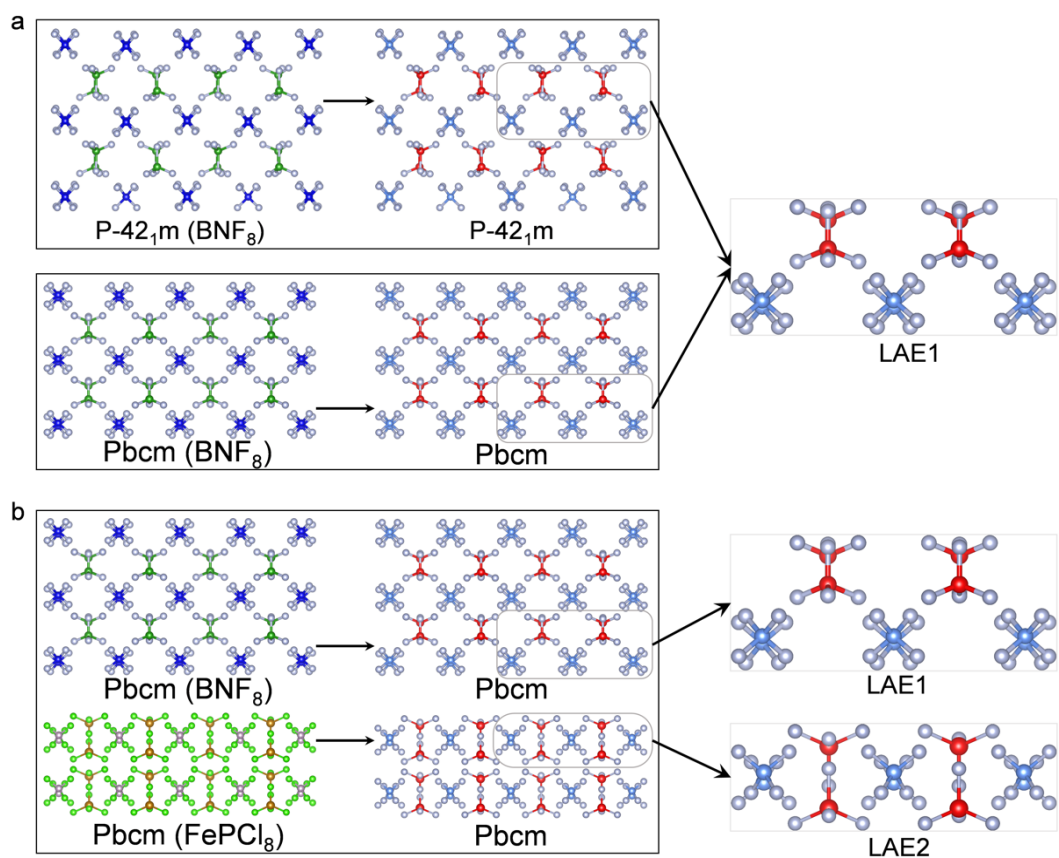

**Figure S1.** (a) Two $BNF_8$ structures with different space groups have the same local atomic environment (LAE), (b) $BNF_8$ structure and $FePCl_8$ structure have the same space group but different LAEs.

- **The "compromised maximum growth rate" method**

As shown in Figure S2, this method first determines whether the number of samples (N) is 1; if so, then the clustering threshold value is set as 1 and the clustering process ends. If N is 2, the only clustering distance ($d_1$) will be compared to the average value of the all maximum clustering distances in each clustering group (AMXD) (or another suitable value). If $d_1$ is less than the AMXD in step 1 while in step 2, it is less than 0.8*AMXD, then the number of clusters will be set to 1; otherwise, the clustering threshold value will be set to 2. If N is greater than 2, the maximum clustering distance ($d_{N-1}$) will be compared to the average value of the all minimum clustering distances of each clustering group. If $d_{N-1}$ is less than this value, then the clustering threshold value will be set to 1. Otherwise, the clustering will be cut off at the maximum growth rate point of the clustering distance. Moreover, to prevent the cut-off point from being found at the bottom of the clustering dendrogram, which may result in an unreasonable clustering threshold value, we set another rule in this method. If the cut-off point is present at the bottom of the clustering dendrogram, it will be reset at the point closest to the average value of the all average clustering distances of each clustering group.

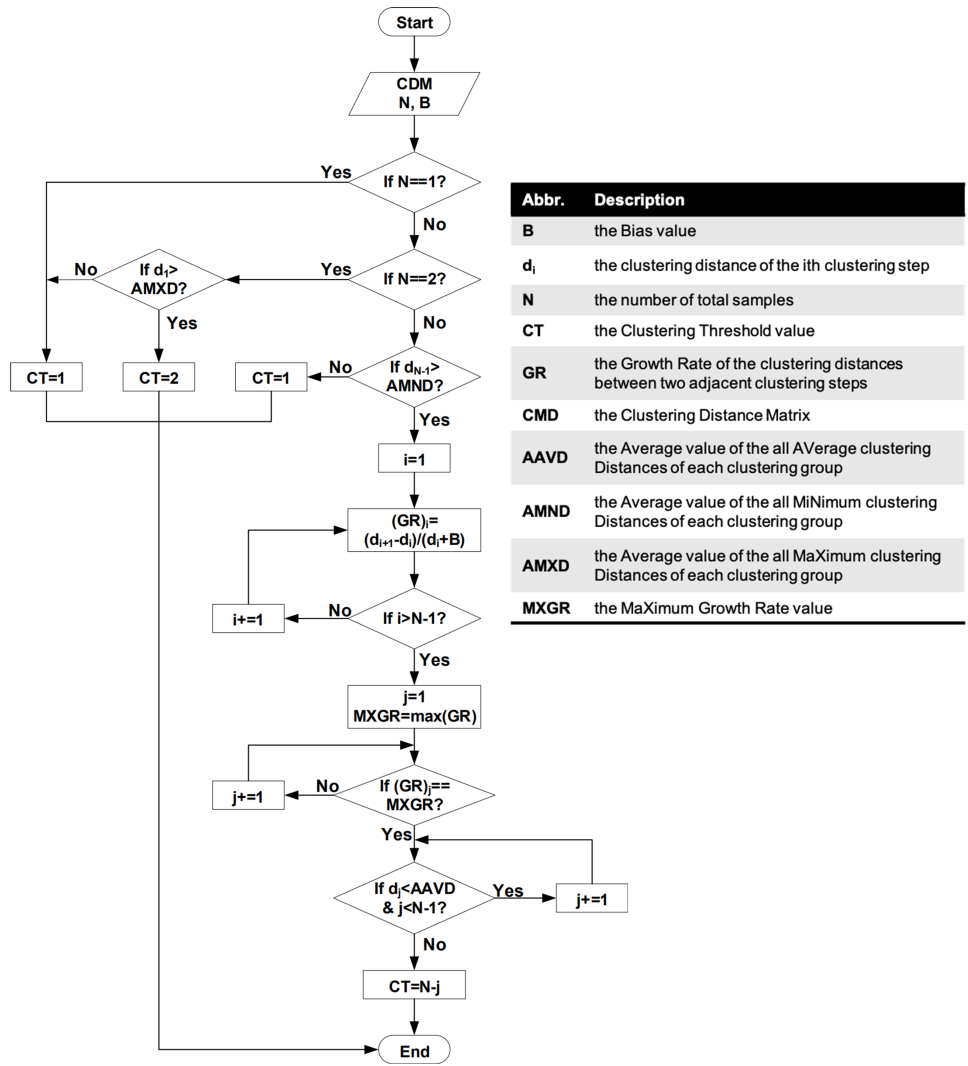

**Figure S2.** Workflow chart of selecting a suitable clustering threshold value. The statistic properties of the entire dataset are taken into consideration to determine a suitable hierarchical clustering threshold value.

- **The clustering results of (A1B1C10)₂ group based on different linkage methods**

The clustering dendrograms of $(A_1B_1C_{10})_2$ group based on different linkage methods are showed in Figure S3-S6. The clustering results based on "single", "complete", "centroid", "ward" linkage are all the same as "average" linkage's clustering results.

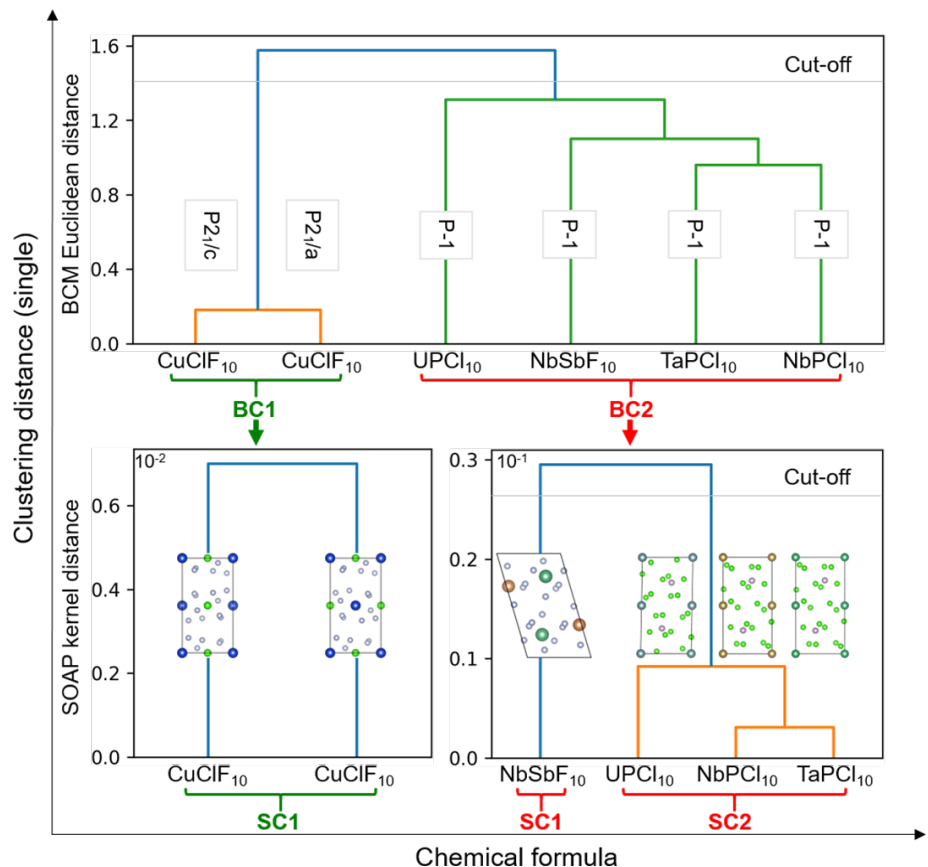

**Figure S3.** The clustering dendrogram of $(A_1B_1C_{10})_2$ group based on "single" linkage.

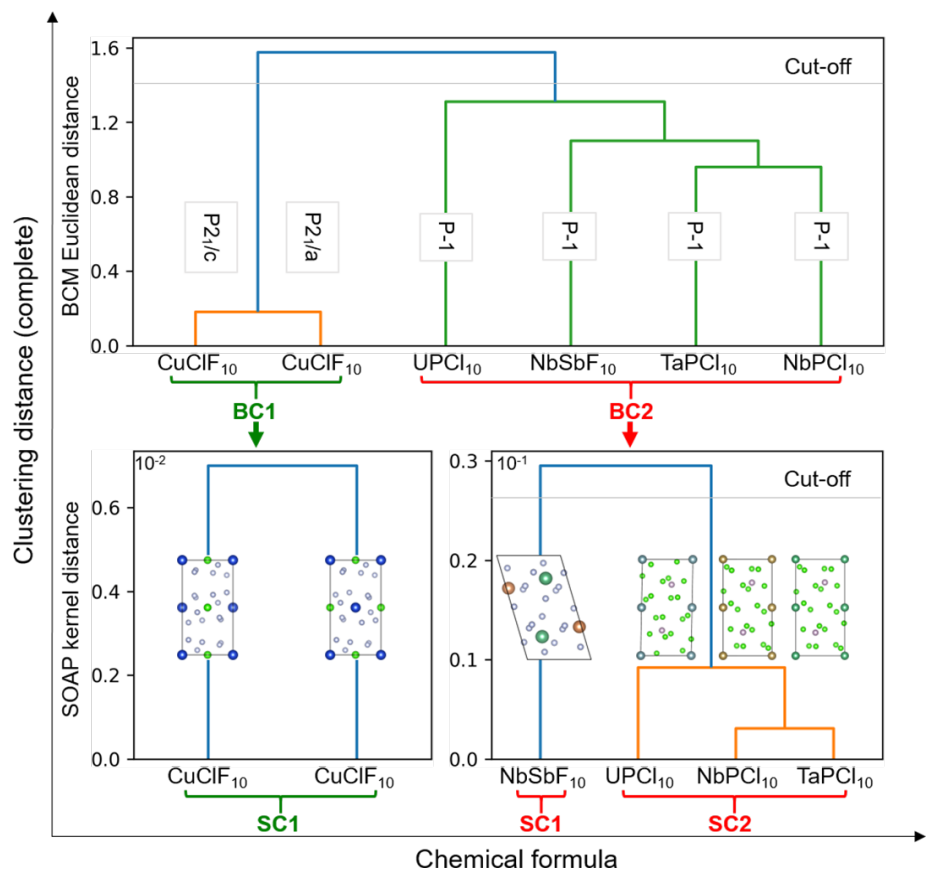

**Figure S4.** The clustering dendrogram of $(A_1B_1C_{10})_2$ group based on "complete" linkage.

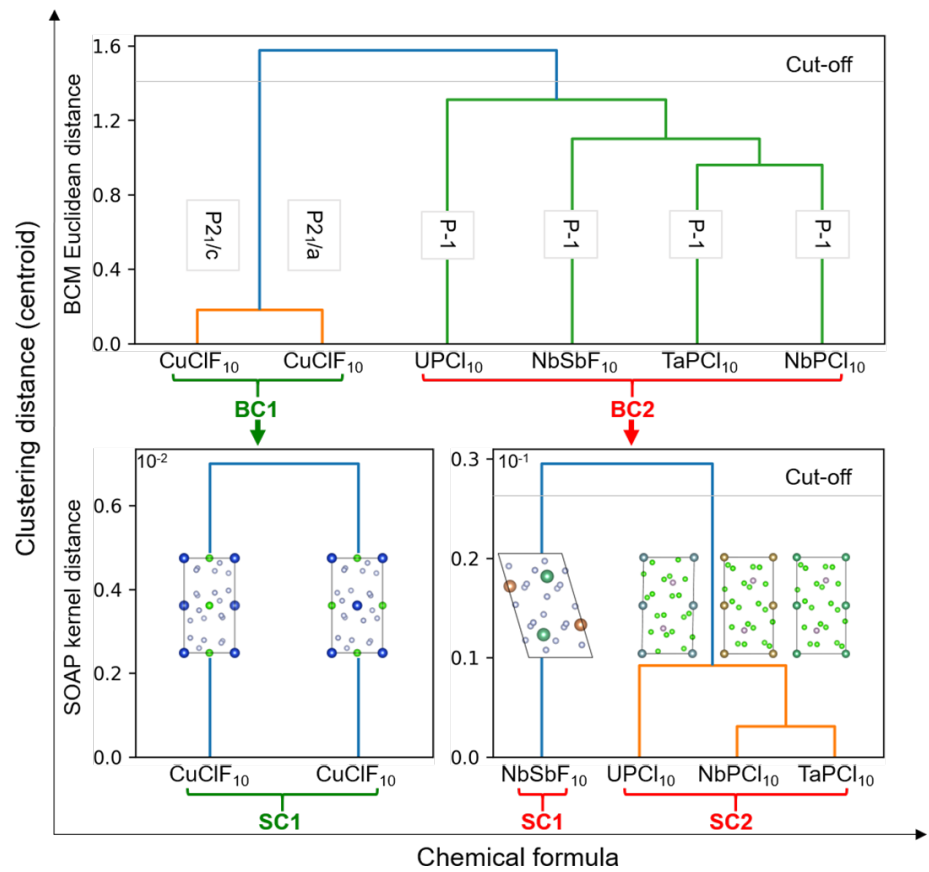

**Figure S5.** The clustering dendrogram of $(A_1B_1C_{10})_2$ group based on "centroid" linkage.

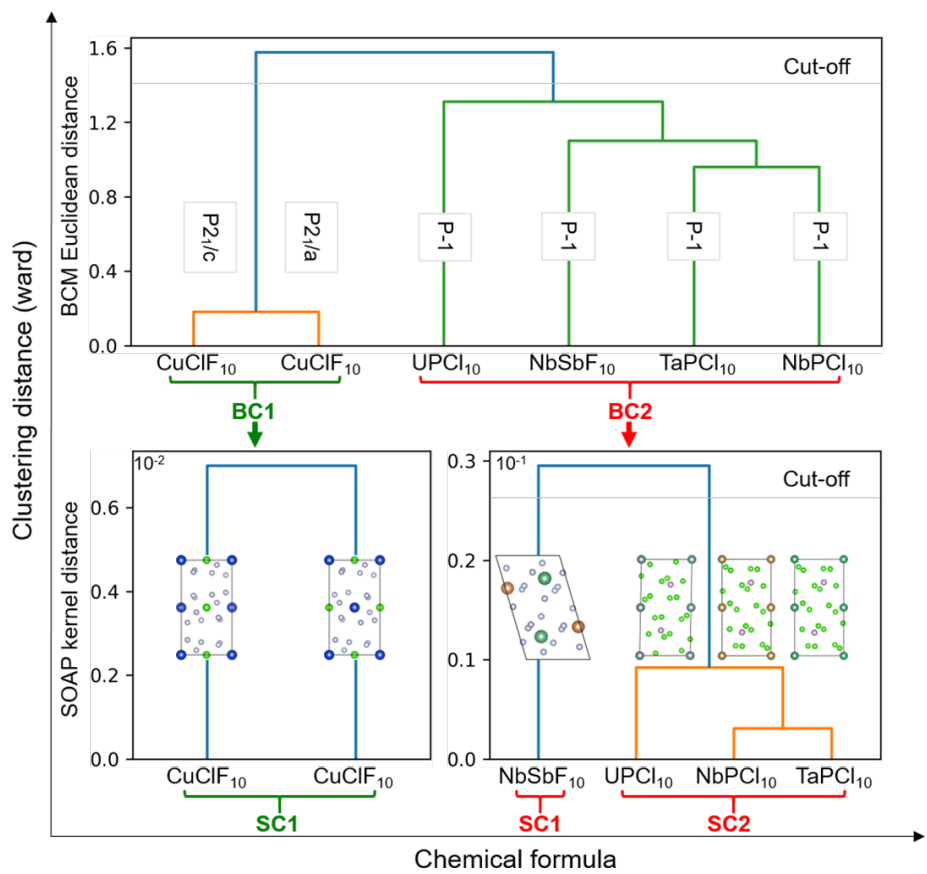

**Figure S6.** The clustering dendrogram of $(A_1B_1C_{10})_2$ group based on "ward" linkage.

- **The correlation between the energy difference and the structural distance for MgCl$_2$ structures**

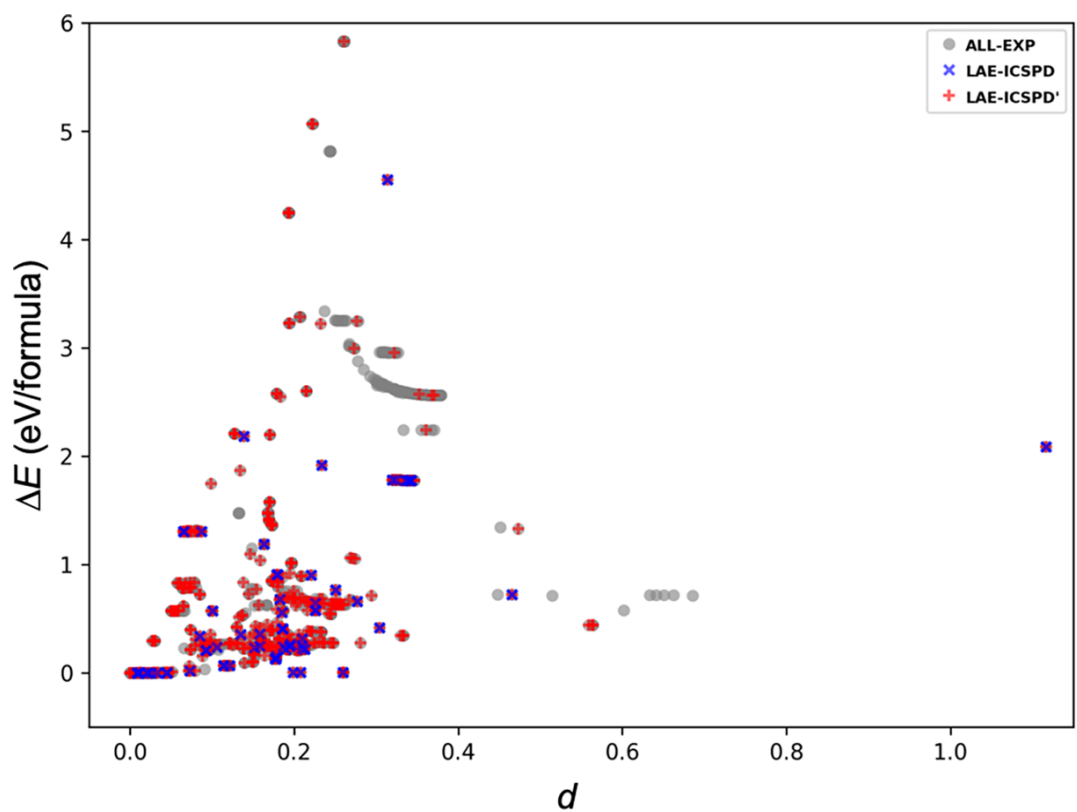

**Figure S7.** The scatter diagram about correlation between the energy difference ($\Delta E$) and the structural distance ($d$) for the experimentally known MgCl$_2$ structure and virtual MgCl$_2$ structures. ALL-EXP: The virtual MgCl$_2$ structures generated based on all experimentally known structures. LAE-ICSPD: The virtual MgCl$_2$ structures generated based on prototype structures stored in our local atomic environment (LAE) based inorganic structure prototype database (ICSPD). LAE-ICSPD`: The virtual MgCl$_2$ structures generated based on prototype structures stored in LAE-ICSPD` database.

- **Supporting references**